\documentclass{svproc}
\usepackage[utf8]{inputenc}

\usepackage{amsmath,amssymb,amsfonts}
\usepackage{graphicx}
\usepackage{siunitx}	
\usepackage[FIGTOPCAP]{subfigure}
\usepackage{float}
\usepackage{verbatim}
\usepackage[font=small]{caption}
\usepackage{xcolor}
\usepackage{cancel}
\usepackage{multirow}

\usepackage{url}

\begin{document}
\mainmatter              
\title{Modelling hydrogen storage in metal hydrides}
\titlerunning{Modelling hydrogen storage}  
%
\author{Francesc Font\inst{1} \and Attila Husar\inst{1} \and Tim Myers\inst{2} \and
Maria Aguareles\inst{3} \and Esther Barrabés\inst{3}}
\authorrunning{Francesc Font et al.} 
%
\tocauthor{Francesc Font, Attila Husar, Tim Myers, Maria Aguareles, Esther Barrabés}
\institute{Department of Fluid Mechanics, Universitat Politècnica de Catalunya - BarcelonaTech, Av. Eduard Maristany, 16, Edifici A, Barcelona 08019, Spain\\
\email{francesc.font@upc.edu},
\and
Centre de Recerca Matemàtica, Campus de Bellaterra, Edifici C, Bellaterra 08193, Barcelona, Spain
\and 
Department of Computer Science, Applied Mathematics and Statistics, Universitat de Girona, Campus de Montilivi, Girona 17071, Catalunya, Spain
}

\maketitle              

\begin{abstract}
We develop a one-dimensional mathematical model for the loading process of hydrogen in a metal hydride tank. The model describes the evolution of the density and pressure of the hydrogen gas, the temperature of the tank, the averaged velocity of the gas through the porous metal structure, and the transformed fraction of metal into a metal hydride. The non-dimensionalisation of the model indicates a possible reduction of the system of equations and also shows that the density and the transformed metal fraction may be decoupled from the temperature equation. The reduced model is solved numerically. Introducing a spatial dependence into the kinetic reaction constant allows to explain unexpected temperature gradients observed in experiments.

\end{abstract}
\section{Introduction}

Motivated by the global warming crisis and the active role of policy makers to reduce CO$_2$ emissions, the general interest in hydrogen as a potential new green energy vector has increased remarkably in the last decade. In line with this, research towards safe and efficient methods of storing hydrogen has also increased. Hydrogen storage in metal hydrides (MH) presents several advantages in terms of safety and storage capacity with respect to classical methods such as compressed or liquefied hydrogen storage \cite{Sch2001}. However, the thermal management of MH tanks is challenging due to the large quantities of heat released during hydrogen loading. High equilibrium pressures and temperatures for the metal to metal hydride reaction are also a challenge for its use in applications operating around room temperature. Thus, more research needs to be done to make MH tanks viable for their implementation in practical applications.   

\begin{figure*}[h]
    \centering
    \includegraphics[width=0.3\textwidth]{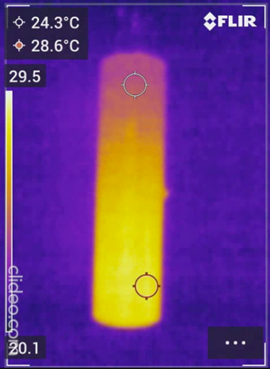}\hspace{0.3cm}\includegraphics[width=0.5\textwidth]{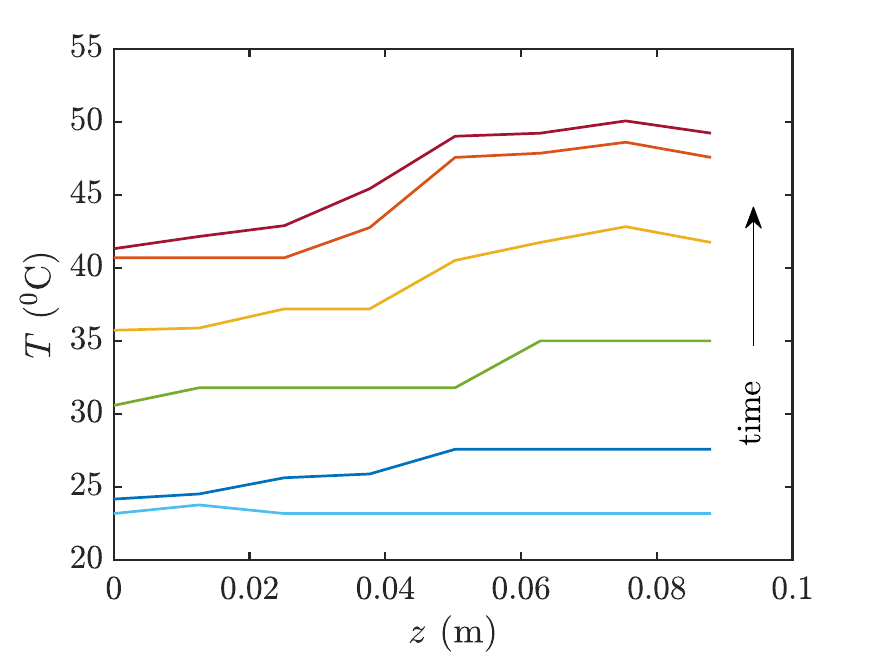}
    \caption{Left: Thermal image of the lab-scale metal hydride hydrogen storage canister Hydrostick PRO during hydrogen loading. Right: Temperature profiles at different times during the first 8 min  of the loading process obtained from the thermal camera images. The bottom/top profiles correspond to the initial/final times.}
    \label{fig_1}
\end{figure*}

In this contribution, we formulate and solve a one-dimensional mathematical model describing the hydrogen loading process in a MH tank. We base our model on experiments performed with the lab-scale hydrogen storage canister Hydrostick PRO from the company Horizon Educational. Typically, the metal alloys used in MH tanks are of type AB2 or AB5 (i.e., alloys formed by 1 atom of component A and 2 or 5 atoms of component B, respectively). The Hydrostick PRO is based in a AB5 metal alloy but the particular composition of the metal is unknown (proprietary information). In these experiments, hydrogen is input at a constant mass flow rate and the temperature evolution of the canister is recorded with a FLIR C5 thermal camera (Fig.~\ref{fig_1}, left). The images are  later postprocessed with Matlab and converted into data points for the temperature along the canister length (see right image in Fig.~\ref{fig_1}). The mathematical model describing the hydrogen loading process in these experiments is presented in Section \ref{model}. The model is then solved numerically. The results are discussed in Section \ref{results} and, finally, the main conclusions are presented in Section \ref{conclusions}.   

\section{One-dimensional mathematical model}
\label{model}
We use the hydrogen storage experiment shown in Figure~\ref{fig_1} as a reference set-up to develop our model. The size and the material properties of the MH canister are described in Table~\ref{fig_1}. Since the particular metal hydride in the Hydrostick PRO is unknown, we choose  values for the material properties which are similar to those found in the literature for AB5 alloys \cite{Sandrock99}. 

Hydrogen is pumped through the top of the tank at a constant mass flow rate and travels from top to bottom through the pore space of the metallic porous bed. When hydrogen  contacts  the boundaries of the porous media the hydrogenation reaction begins; the hydrogen is absorbed by the metal which transforms into a metal hydride. The hydrogenation reaction liberates large quantities of heat, rapidly increasing the temperature of the tank.

To formulate a one-dimensional model we consider a cylindrical tank of radius $r$ and length $L$. We assume radial symmetry and consider the radial component of the velocity field to be negligible. Thus, conservation of mass in the gas phase leads to 
\begin{equation}\label{density}
	\varepsilon \frac{\partial \rho}{\partial t} + \varepsilon \frac{\partial}{\partial z} \left(\rho u \right)  = -\rho_m\, {\omega}\, (1-\varepsilon) \frac{\partial \alpha}{\partial t}\,,  
\end{equation}
where $\rho$ is the density of hydrogen, $u$ the average interstitial velocity in the axial direction, $\varepsilon$ (non-dimensional) the porosity, $\rho_m$ the density of the metal hydride, $\omega$ the maximum mass fraction of hydrogen in the metal and $\alpha$ the fraction of metal transformed into a metal hydride. The transformation of metal into a metal hydride, follows the kinetic law  
\begin{equation}\label{reaction}
	\frac{\partial \alpha}{\partial t} = k \left(\frac{p-p_\textrm{eq}}{p_\textrm{eq}}\right) (1-\alpha)\,,    
\end{equation}
where $k$ is the reaction constant and $p_\textrm{eq}$ the equilibrium pressure at which the reaction occurs. Describing the kinetics of hydrogen absorption into the metal via the linear driving force \eqref{reaction} is a standard approach in the literature \cite{Chaise2010,Bho2017}. Therefore, we also take the linear driving force approach in the current work. Note that in the most general case, $k$ depends on temperature via an Arrhenius law and $p_\textrm{eq}$ depends on temperature via a van't Hoff relation \cite{Chaise2010,Lutz2019,Chaise2009}. As a first approximation, we consider $k$ and $p_\textrm{eq}$ constant.  

The governing equation for the evolution of the temperature in the tank is  
\begin{equation} \label{temperature}
\begin{split}
    	(1-\varepsilon)\rho_m c_{p,m}\frac{\partial T}{\partial t} &+ \varepsilon\rho c_{p,g}\frac{\partial (Tu)}{\partial z} = \lambda \frac{\partial^2 T}{\partial z^2} \\
     &+ \frac{2h_a}{r}(T_0-T)+\frac{\rho_m \omega (1-\varepsilon)\Delta H}{M}\frac{\partial \alpha}{\partial t}\,,
\end{split}
\end{equation}
where $T$ is the averaged temperature over the cross sections of the tank, $\Delta H$ the enthalpy of the reaction, $M$ (kg/mol) the molar mass of hydrogen, $\lambda$ an effective thermal conductivity, $h_a$ a heat transfer coefficient between the tank and the surrounding air, $r$ the radius of the tank, $T_0$ the ambient room temperature, and $c_{p,g}$, $c_{p,m}$ the specific heats of hydrogen and metal, respectively. Note in writing equation \eqref{temperature} we have  assumed thermal equilibrium between the hydrogen and the metal and  have used the approximation $\varepsilon \rho c_{p,g} + (1-\varepsilon) \rho_m c_{p,m}\approx (1-\varepsilon) \rho_m c_{p,m}$ to obtain the first term of \eqref{temperature}. 

The average velocity of the gas in the axial direction of the tank is described by  Darcy's law 
\begin{equation}\label{vChaise}
	\frac{\partial p}{\partial z} = -\beta  u
\end{equation}
where $\beta = \mu \varepsilon/K$, with $\mu$ the viscosity of hydrogen and $K$ the permeability of the porous media. Finally, pressure, density and temperature, are related via the ideal gas law 
\begin{equation}\label{idealgas}
	p = \rho R' T
\end{equation}
where $R'=R/M$ with $R$ the universal ideal gas constant. We note that the mathematical model \eqref{density}-\eqref{idealgas} holds strong similarities with those describing contaminant removal in adsorption columns \cite{Myers2020a,Myers2020b,Sips}. They consist of radially averaged 1D models of a gas (or liquid) phase flowing through a porous media, with mass transfer between the gas and the porous media modelled via a linear driving force or more elaborate forms.  

Hydrogen enters the tank at a constant mass flow rate and it does not escape the tank, thus
\begin{equation}
	(\rho u)|_{z=0} = \frac{\dot{m}}{A}\, ,\qquad (\rho u)|_{z=L} =0\, ,
\end{equation}
where $\dot{m}A^{-1}$ is a prescribed constant value (see Table~\ref{tab_param}). Note that $z=0$ represents the inlet or top of the tank while $z=L$ represents the end or bottom of the tank. For the temperature 
\begin{equation}
	-\lambda\left.\frac{\partial T}{\partial z}\right|_{z=0} = h  \left(T_0-T|_{z=0}\right)\,, \qquad -\lambda\left.\frac{\partial T}{\partial z}\right|_{z=L} = h_a \left(T|_{z=L}-T_0\right)\,,
\end{equation}
where $h$ is an effective heat transfer coefficient, involving natural convection with air and some forced convection  due to hydrogen flow at the tank inlet. Finally, the pressure and velocity of the gas in the tank are subject to
\begin{equation*}
\left.	u\right|_{z=L} = 0\,, \qquad \left.\frac{\partial p}{\partial z}\right|_{z=L}=0.
\end{equation*}

The initial conditions are 
\begin{equation*}
	p(z,0) = p_\textrm{eq},\quad T(z,0) = T_0,\quad \rho(z,0) = \rho_0,\quad \alpha(z,0) = \alpha_0,\quad u(z,0) = u_0\,, 
\end{equation*}
where $T_0$ is the ambient room temperature, $\rho_0 = p_\textrm{eq}/(R'T_0)$ and $u_0 = \dot{m}/(A\rho_0)$. The initial reacted fraction, $\alpha_0$, represents a small quantity of metal in the form of metal hydride before hydrogen loading initiates. 

\begin{table}[h]
	\centering
	\begin{tabular}{|c|c|c|}
		\hline 
		parameter & description & value \\
		\hline
		$\rho_m$      & metal density  &  8000 kg/m$^3$\\
		$c_{p,m}$     & metal specific heat & 400 J/kg\,K \\
		$c_{p,g}$     & gas specific heat &  14000 J/kg\,K \\ 
		$\lambda$     & effective thermal conductivity &  1 W/m\,K\\ 
		$\varepsilon$ & porosity  & 0.45 \\ 
		$\omega$      & max. mass fraction of $H_2$ in metal & 0.01 \\
		$M$           & molar mass of $H_2$ & 0.002 kg/mol \\ 
		$\Delta H$    & molar enthalpy of reaction & 30000 J/mol\\
		$p_\textrm{eq}$      & equilibrium pressure for reaction & 2 bar\\
		$\mu$         & viscosity of $H_2$ & 8.9$\cdot10^{-6}$ Pa\,s \\
		$K$           & permeability of porous structure  & 5.75$\cdot10^{-14}$ m$^2$\\  
		$L$           & height of the tank & 8.8$\cdot10^{-2}$ m\\  
		$r$           & radius of the tank & 1.1 $\cdot10^{-2}$ m\\  
		$A$           & cross-section of inlet region  & 9.5$\cdot10^{-5}$ m$^2$ \\
		$\dot{m}$     & mass flux through inlet region & 5.3$\cdot10^{-7}$  kg/s \\
		$k$         & kinetic constant & $2\cdot10^{-4}$ s$^{-1}$\\
				$T_0$         & ambient temperature & 23\,$^o$C\\
  				$h_a$         & air-canister heat transfer coef. & 25\,W/m$^2$K\\
  				$h$         & inlet heat transfer coef. & 100\,W/m$^2$K\\
		\hline
	\end{tabular}
	\caption{Summary of parameter values for the model. The specific metal properties are unknown and are chosen to be consistent with those of an AB5 alloy. The indicated geometry and experimental conditions come from laboratory measurements.}
	\label{tab_param}
\end{table}

\subsection{Non-dimensionalisation} 

We non-dimensionalise the model with the dimensionless variables 
\begin{equation}
	\hat{t} = \frac{t}{\tau}\,,\quad\hat{z} = \frac{z}{\mathcal{L}}\,,\quad \hat{\rho} = \frac{\rho}{\Delta \rho}\,,\quad \hat{p} = \frac{p-p_{eq}}{\Delta p}\,,\quad \hat{T} = \frac{T-T_0}{\Delta T}\,,\quad\hat{u} = \frac{u}{\mathcal{U}}\,,
\end{equation}
and choose the scales $\tau=1/k$, $\mathcal{L} = L$, $\Delta \rho = \rho_0$, $\Delta p = p_\textrm{eq}$ and $\mathcal{U} = u_0$. Since heat is primarily lost by cooling at the boundaries and generated by the reaction, the temperature scale $\Delta T = \rho_m\,(1-\varepsilon)\omega \Delta H r/(2h_aM\tau)$ is chosen to balance the penultimate and final terms in the energy equation. The hydriding fraction $\alpha$ is already a dimensionless variable. 

Substituting the dimensionless variables in the governing equations and dropping the hat notation, we obtain 
\begin{subequations}
    \begin{align}
	\delta_1\,\frac{\partial \rho}{\partial t} + \delta_2\, \frac{\partial}{\partial z}(\rho u) &= - \frac{\partial \alpha}{\partial t}\,,\label{mass_non}\\
	\delta_4\, \frac{\partial T}{\partial t} + \delta_5\, \rho\, \frac{\partial}{\partial z}\left[\left(1+\delta_3 T \right)u\right] &= \delta_6\, \frac{\partial^2 T}{\partial z^2} - T+\frac{\partial \alpha}{\partial t} \,,\label{temp_non}\\
	\frac{\partial \alpha}{\partial t} &=  p \left(1-\alpha\right)\,,\label{reaction_non}\\   
	\frac{\partial p}{\partial z} &= -\delta_7\, u\,, \label{Darcy_non}\\
	1 + p &= \rho \left(1+\delta_3\, T\right)\,, \label{ideal_gas_non}
\end{align}
\end{subequations}
where 
\begin{equation}
    \begin{split}
        &\delta_1 = \frac{\varepsilon \Delta \rho }{\rho_m (1-\varepsilon)\,\omega } \,,\quad \delta_2 = \frac{\tau \varepsilon \Delta \rho \, \mathcal{U}}{\rho_m (1-\epsilon)\,\omega\, \mathcal{L}} \,,\quad \delta_3 = \frac{\Delta T}{T_0} \,,\quad \delta_6 = \frac{\lambda r}{2 h_a\, \mathcal{L}^2} \,, \\ 
	&\delta_4 = \frac{\rho_m (1-\varepsilon) \,c_{p,m}\, r }{2 h_a\, \tau } \,, \quad \delta_5 = \frac{\varepsilon \Delta \rho \, c_{p,g}\, \mathcal{U}\, T_0\, r}{2 h_a\, \Delta T \,\mathcal{L}} \,,\quad \delta_7 = \frac{\beta\, \mathcal{U}\,\mathcal{L}}{\Delta p} \,.\ 
    \end{split}
\end{equation}
The boundary conditions reduce to 
\begin{subequations}
\label{bc}
    \begin{align}\label{bc_nonA}
    &(\rho u)|_{z=0} = 1\,, \quad \quad u|_{z=1} = 0\, ,\\
    \label{bc_nonB}
&\text{Bi}_1^{-1}\left.\frac{\partial T}{\partial z}\right|_{z=0} =\, T|_{z=0} \,,\quad  -\text{Bi}^{-1}_2\left.\frac{\partial T}{\partial z}\right|_{z=1} =  \, T|_{z=1}  \,,
\end{align}
\end{subequations}
where $\text{Bi}^{-1}_1 = \lambda/(\mathcal{L} h)$ and $\text{Bi}^{-1}_2= \lambda/(\mathcal{L} h_a)$, and the initial conditions are
\begin{equation} \label{ic_non}
	T(z,0) = 0\,, \quad \rho(z,0) = 1 \,,\quad p(z,0) = 0\,, \quad u(z,0) = 1\,,\quad \alpha(z,0)=\alpha_0\,. 
\end{equation}

\subsection{Reduced model} 



The parameter values in Table \ref{tab_param} show that $\delta_3 = 0.0973 \ll 1$. Neglecting these terms in \eqref{temp_non} and \eqref{ideal_gas_non}  decouples the variables $\rho$, $p$, $u$, $\alpha$ from $T$. Physically this implies a weak dependence on temperature. Within this approximation, $p\approx\rho-1$ and the velocity in Darcy's law can be expressed as a function of $\rho$. Then, the system of equations \eqref{mass_non}-\eqref{ideal_gas_non} reduces to
\begin{subequations}
\label{red}
    \begin{align}
	&\delta_1\,\frac{\partial \rho}{\partial t} - \frac{\delta_2}{\delta_7} \, \frac{\partial}{\partial z}\left(\rho \frac{\partial \rho}{\partial z} \right) = - \frac{\partial \alpha}{\partial t}\,,\label{density_red}\\
	&\delta_4\, \frac{\partial T}{\partial t} - \frac{\delta_5}{\delta_7}\, \rho\, \frac{\partial^2 \rho}{\partial z^2} = \delta_6\, \frac{\partial^2 T}{\partial z^2} - T+\frac{\partial \alpha}{\partial t} \,,\label{temp_red}\\
	&	\frac{\partial \alpha}{\partial t} =  \left(\rho -1\right) \left(1-\alpha\right)\,. \label{reac_red}
\end{align}
Once $\rho$, $\alpha$, $T$ are found, $u$ and $p$ can be obtained from \eqref{Darcy_non}-\eqref{ideal_gas_non} (neglecting terms with $\delta_3$) .  

The boundary conditions for the temperature are \eqref{bc_nonB} 
and the resulting boundary conditions for the density are 
\begin{equation}
	\left.\left(\rho \frac{\partial \rho}{\partial z}\right)\right|_{z=0} =- \delta_7 \,, \quad \left.\frac{\partial \rho}{\partial z}\right|_{z=1} = 0\,.
\end{equation}
\end{subequations}
We assume that the equation for $\alpha$ holds on the boundaries of the domain, so no boundary conditions for $\alpha$ are required. Finally, the initial conditions are those specified in \eqref{ic_non}.

In the following section we discuss the results obtained by numerically solving the reduced problem \eqref{red}.

\section{Results and discussion}
\label{results}

The system \eqref{density_red}-\eqref{reac_red} is solved numerically by finite differences, using an explicit time discretisation and central differences in space. The numerical scheme is implemented in Matlab. The time step is appropriately chosen to fulfill the stability requirements of the diffusion terms in (\ref{density_red}, \ref{temp_red}) \cite{NRecip}.  

\begin{figure*}[h]
	\centering
	\includegraphics[width=0.5\textwidth]{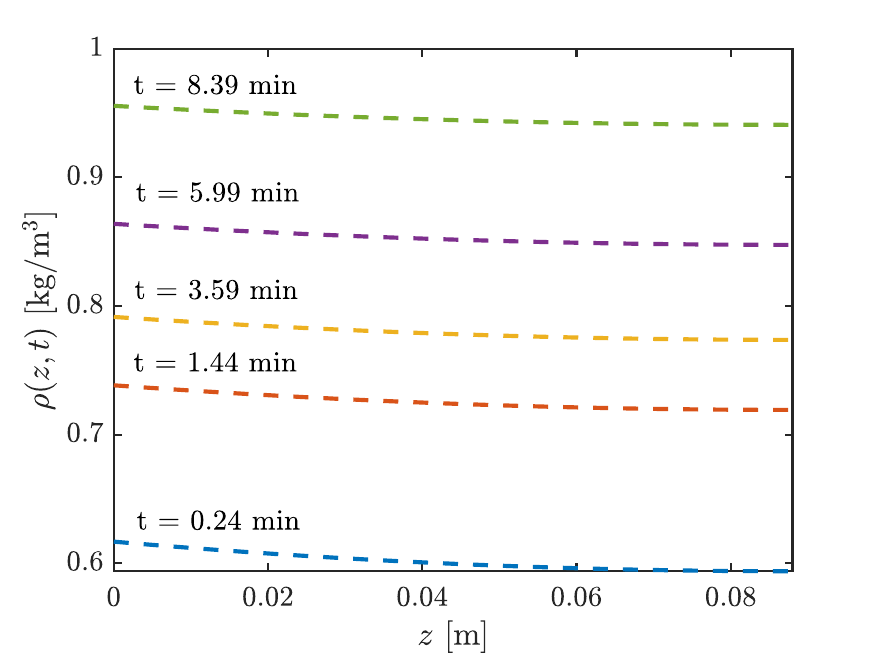}\includegraphics[width=0.5\textwidth]{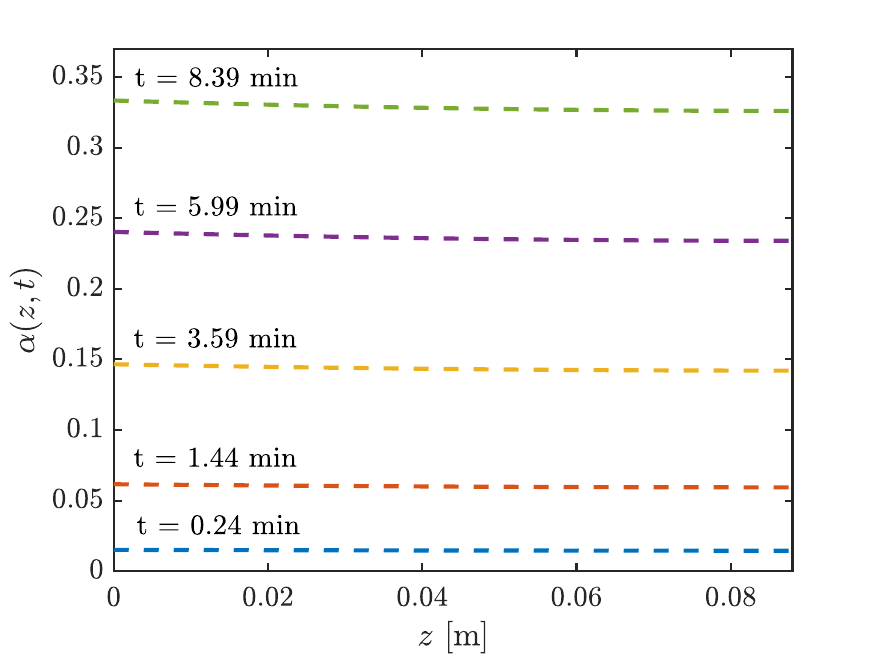}
	\caption{Density (left panel) and reacted fraction (right panel) profiles at different times obtained by solving \eqref{red} with constant $k$ during hydrogen loading. }
	\label{fig_2}
\end{figure*}

On the left panel of Figure \ref{fig_2}, we show the evolution of the density of hydrogen along the tank at several times during the loading process. Density throughout the tank is distributed almost homogeneously with   relatively flat profiles along the $z$ axis. As expected, density is highest at the inlet where hydrogen is being loaded. Similarly, on the right panel, the reacted fraction shows very  small gradients throughout the tank, being also higher at the inlet. This is consistent with the fact that spatial gradients in the governing equation for $\alpha$ are only introduced through variations in $\rho$.        

\begin{figure*}[h]
	\centering
	\includegraphics[width=0.5\textwidth]{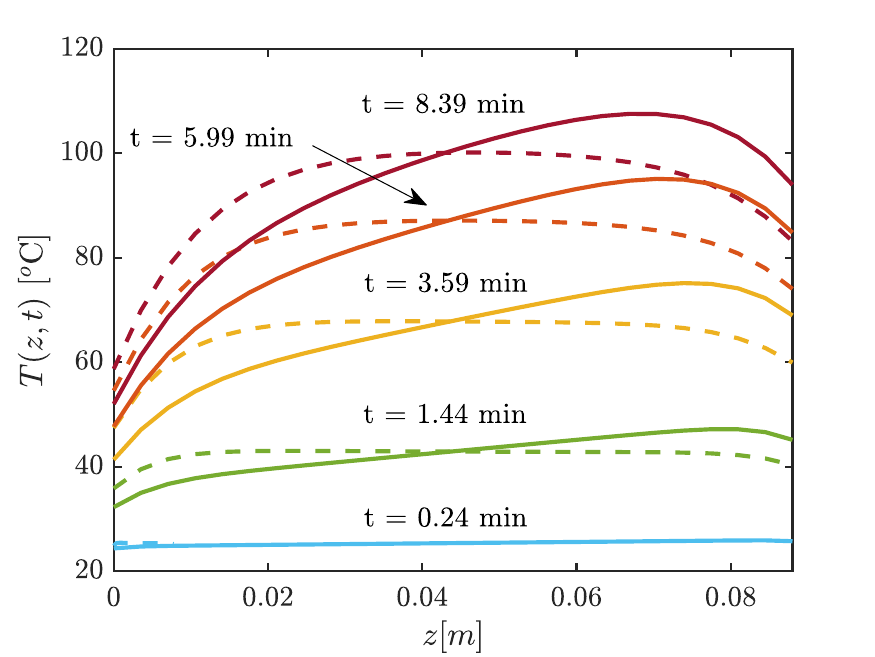}\includegraphics[width=0.5\textwidth]{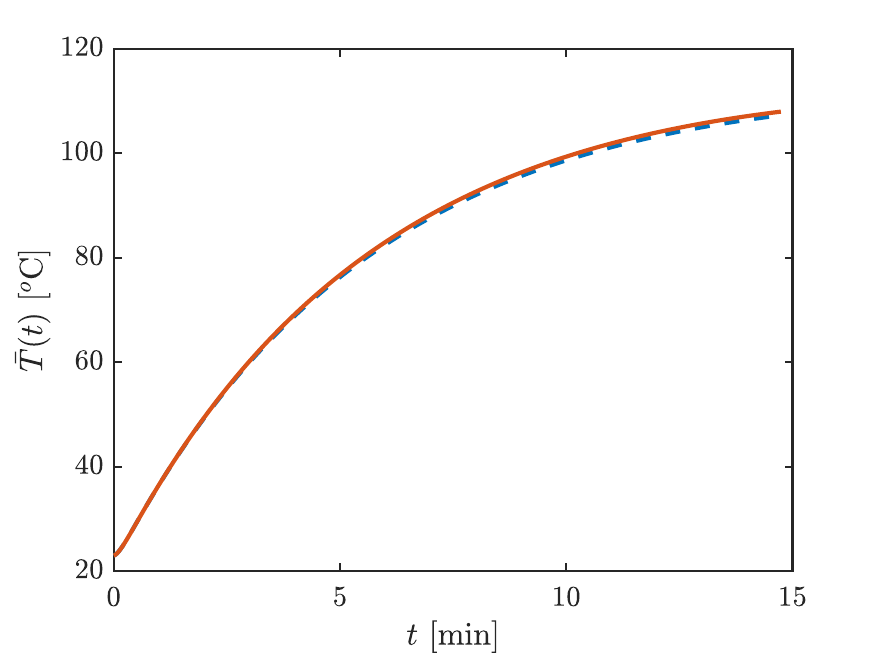}
	\caption{Left panel: Temperature profiles at different times during hydrogen loading obtained by solving \eqref{red} with constant and variable $k$. Right panel: Evolution the spatially averaged temperature across the entire tank. In both panels, the dashed and solid lines indicate the cases with constant and variable $k$, respectively.}
	\label{fig_3}
\end{figure*}

The dashed lines in the left panel of Figure \ref{fig_3} are the temperature profiles corresponding to the simulation of Figure \ref{fig_2}. The temperature in the tank tends to increase the loading proceeds, due to the heat liberated from the hydriding reaction. The temperature increase is lower at the inlet than at the bottom of the tank. This is due to the convective cooling which is greater at the open inlet than the sealed bottom. However, we note that the temperature gradient is small in the central region, in contrast to the experimental profile of Figure \ref{fig_1}. The difference seems unlikely to come from the uncoupling of the temperature, since the neglected term suggests errors of the order 10\%.   As a first attempt to reconcile the differences we consider a new effect in the model, namely degradation of the metal.

Here we consider the effect of a spatially-variable kinetic coefficient, $\bar{k}=k f(z)$. The physically justification lies in the impurities 
carried inside the tank by the hydrogen gas (for example oxygen). These impurities would primarily react near the inlet resulting in a degradation of the metal there, with a resultant loss of storage capacity of the metal. This suggests  modifying the metal transformation rate, equation  \eqref{reaction_non}, to 
\begin{equation}
    \frac{\partial\alpha}{\partial t} = f(z)(1-\alpha)\, .
\end{equation}
As a first approximation we take a linear form for $f(z)$ 
\begin{equation}\label{spatial_k}
	\bar{k}(z) = k f(z)=k\left(\frac{2}{3}\left(1+\frac{z}{L}\right)\right) \, .
\end{equation}
The value 2/3 is speculative, with the simple goal of reducing $k$ at the inlet to below its previous value, as shown in Table \ref{tab_param} (so reflecting a degradation effect). The linear form allows for the reaction to be increasingly strong along the tank axial direction, thereby releasing more heat near the bottom of the tank.  

The solid lines in the left panel of Figure \ref{fig_3} show the temperature profiles from the numerical solution of the model using the spatially-varying form of $k$ \eqref{spatial_k}. When compared with the profiles for a constant $k$, the ones with variable $k$ seem to follow a trend  similar to the experimental ones.  Finally, the evolution of the spatially averaged temperature across the entire tank, $\bar{T}$, is shown in the right panel of Figure \ref{fig_3}. The average temperature in the two cases considered, constant and spatially-variable $k$, evolves almost identically, ensuring that the energy released during the process is the same in both cases.  

\section{Conclusions}
\label{conclusions}
This preliminary mathematical model appears to provide a reasonable approximation to the  experimental data. However the observed temperature rise near the middle of the tank is not captured. It appears that the constant reaction rates used in the model are not sufficiently strong to generate this rise. Given that the particular metal hydride of the experiment considered in this work is unknown, we proposed a simple linear profile for the reaction rate, which mimics the degradation of material close to the inlet. 
The temperature gradients obtained using the linear profile for the reaction rate show much better agreement with the experimental ones than those obtained using the constant reaction rate.

\section*{Acknowledgements}

This publication is part of the research projects PID2020-115023RB-I00 and TED2021-131455A-I00 (funding T. Myers, M. Aguareles, E. Barrabés and F. Font) financed by
MCIN/AEI/ 10.13039/501100011033/, by “ERDF A way of making Europe” and by “European Union NextGenerationEU/PRTR”. F. Font is a Serra-Hunter fellow from the Serra-Hunter Programme of the Generalitat de Catalunya. T. Myers acknowledges the CERCA Programme of the Generalitat de Catalunya. The work was also supported by the Spanish State Research Agency, through the Severo Ochoa and Maria de Maeztu Program for Centres and Units of Excellence in R\&D (CEX2020-001084-M). 

%
%

\end{document}